\documentclass[11pt,a4paper,portrait,times]{article}
\usepackage[english]{babel}
\usepackage{url,hyperref,lineno,microtype,subcaption}
\usepackage{longtable}
\usepackage[pdftex]{graphicx} %incluir figuras
\usepackage[numbers]{natbib}
\usepackage{float}
\usepackage[retainorgcmds]{IEEEtrantools}
\usepackage{soul}
\usepackage{amsmath}
\usepackage{eqnarray}
\usepackage{amssymb}
\usepackage{url} %this package should fix any errors with URLs in refs.
\usepackage{lscape}
\usepackage{pdflscape}
\usepackage[utf8]{inputenc}
\usepackage[english]{babel}
\usepackage{multicol}
\usepackage{xcolor}
\usepackage{wrapfig}
\usepackage[T1]{fontenc}
\usepackage{tgbonum}
\usepackage{rotating}
\usepackage{multirow}

\usepackage[letterpaper,top=2cm,bottom=2cm,left=3cm,right=3cm,marginparwidth=1.75cm]{geometry}

\usepackage{amsmath}
\usepackage{graphicx}

\title{Different Transient Phenomena at the Edges of Traveling Foreshocks}
\author{Primo\v{z} Kajdi\v{c}\,$^{1,*}$, Xóchitl Blanco-Cano\,$^{1}$, Diana Rojas-Castllo\,$^{1}$ and , Nojan Omidi$^{2}$.}

\begin{document}
\maketitle

\begin{enumerate}
\item[$^{1}$] Departamento de Ciencias Espaciales, Instituto de Geof\' isica, Universidad Nacional Aut\'onoma de M\'exico, Mexico City, Mexico
\item[$^{2}$] Solana Scientific Inc., Solana Beach, CA,USA
\end{enumerate}

\begin{abstract}
Past kinetic simulations and spacecraft observations have shown that traveling foreshocks (TFs) are bounded by either foreshock compressional boundaries (FCBs) or foreshock bubbles (FBs). Here we present four TFs with a different kind of structure appearing at one of their edges. Two of them, observed by the Cluster mission, are bounded by a hot flow anomaly (HFA). In one case, the HFA was observed only by the spacecraft closest to the bow shock, while the other three probes observed an FCB. In addition, two other TFs were observed by the MMS spacecraft to be delimited by a structure that we call HFA-like FCB. In the spacecraft data, these structures present signatures similar to those of HFAs: dips in magnetic field magnitude and solar wind density, decelerated and deflected plasma flow and increased temperature. However, a detailed inspection of these events reveals the absence of heating of the SW beam. Instead, the beam almost disappears inside these events and the plasma moments are strongly influenced by the suprathermal particles. We suggest that HFA-like FCBs are related to the evolution and structure of the directional discontinuities of the interplanetary magnetic field whose thickness is larger than the gyroradious of suprathermal ions. We also show that individual TFs may appear together with several different types of transient upstream mesoscale structures, which brings up a question about their combined effect on regions downstream of the bow shock.
\end{abstract}

\section{Introduction}
The region upstream of the Earth's bow shock is highly dynamic, hosting different phenomena. Among the former is our planet's foreshock \citep{eastwood:2005c} that forms upstream of the quasi-parallel section of the bow shock, where the angle between the upstream interplanetary magnetic field (IMF) and the local normal to the shock, $\theta_{BN}$, is less than 45$^{\circ}$. In the foreshock, the pristine solar wind (SW) particles coexist with suprathermal ion populations that are leaked or reflected and energized by the bow shock. Depending on the location inside the foreshock, the suprathermal ions can appear in the form of field-aligned ion beams \citep[FAB, e.q.][]{fairfield:1969, paschmann:1980, meziane:2005}, gyrating population \citep{argo:1967, fuselier:1995, meziane:1998} and diffuse ions \citep{paschmann:1981a, bonifazi:1981, bonifazi:1981b, winske:1984}. The interaction between the SW and the suprathermal ions leads to plasma and magnetic field instabilities and consequently to the formation of ultra-low frequency (ULF) waves. The most studied among them are the so called 30-second waves \citep{eastwood:2005a, eastwood:2005b}, which start out as quasi-monochromatic, transverse waves that are convected towards the bow shock by the SW. As they approach the bow shock, the ULF waves steepen and become more compressive \citep{hada:1987, omidi:1990, scholer:1993a, greenstadt:1995}.

The foreshock is often observed to be delimited by a structure called the foreshock compressional boundary \citep[FCB, e.g.][]{sibeck:2008, omidi:2013b, rojascastillo:2013}. FCBs separate the section of the foreshock populated by compressive ULF waves from either the pristine SW or from the region of the foreshock populated by FABs, but not by the ULF waves. Their signatures in the spacecraft data include a concurrent rise in the magnetic field magnitude and plasma density on the magnetically quiet side, followed by their depletion on the turbulent side. It has been shown in the past that FCBs affect regions downstream of the bow shock: they have been observed to cause local distortions of the magnetopause and transient magnetic field and plasma density perturbations in the magnetosphere \citep{hartinger:2013}.

Past spacecraft observations have revealed the presence of structures called foreshock cavities (FC) that form as a result of a small-scale magnetic flux tube of IMF lines connecting to a nominally quasi-perpendicular ($\theta_{BN}\geq$45$^\circ$) section of the bow shock in a quasi-parallel manner \citep[FC, e.g.][]{sibeck:2002, billingham:2008, billingham:2011}. According to \cite{sibeck:2008}, some FC signatures in the spacecraft data may appear due to a finite period in-and-out encounter of the spacecraft with the foreshock. To distinguish between the two possibilities \cite{omidi:2013} and \cite{kajdic:2017b} refer to FCs associated with flux tubes as traveling foreshocks (TF). In the spacecraft data, TFs exhibit all the signatures of a ``normal'' foreshock, however, they appear bounded by two IMF directional discontinuities, indicating crossing of the spacecraft from one magnetic flux tube into another. TFs have been shown to drive magnetospheric Pc1 and EMIC waves \citep{suvorova:2019}, cause transient compression in the dayside magnetosphere and are a source of magnetosheath jets in the quasi-perpedicular magnetosheath \citep{sibeck:2021, kajdic:2021b}.

Another type of structure commonly observed upstream of the bow shock are hot flow anomalies \citep[HFA,][]{fuselier:1987, paschmann:1988, schwartz:2000, omidi:2007}. These occur when an IMF directional discontinuity intersects the bow shock and the convection electric field ($-{\bf V}\times{\bf B}$) points towards the discontinuity on at least one side. Then the ions reflected by the bow shock are channeled towards upstream along the discontinuity, heating the plasma and making it expand. HFAs can be observed upstream of a quasi-perpendicular bow shock as well as inside the foreshock.

In the spacecraft data, the HFAs exhibit a hot core inside which the B-magnitude and plasma density are strongly depleted and the plasma flow is decelerated and deflected. The core is delimited by leading and trailing edges on at least one side in which magnetic field strength and plasma density are enhanced. HFAs have been shown to cause the magnetopause displacement and magnetospheric perturbations \citep{shen:2018, wang:2018} and their effects on the goemagnetic field have been detected by ground based observatories \cite[e.g.,][]{sibeck:1999, jacobsen:2009, fillingim:2011, kajdic:2024}.

In addition to HFAs, the presence of RDs and TDs in the solar wind and their interaction with the backstreaming ions in the foreshock can result in the generation of other transient phenomena, such as foreshock bubbles \citep[FB,][]{omidi:2010, omidi:2020, omidi:2020b, omidi:2021}. In the hybrid simulations of \cite{omidi:2013} it was shown that depending on plasma conditions, the edge of a TF may be associated with an FB instead of an FCB. Spacecraft observations at Mars
by \citep{madanian:2023} show an example of a traveling foreshock with an FB at one edge.

Foreshock compressional boundaries, traveling foreshocks, hot flow anomalies and foreshock bubbles are some of the types of transient upstream mesoscale structures \citep[TUMS, ][]{kajdic:2024} that are commonly observed upstream of the bow-shock of Earth. Their typical scale sizes range form $\sim$2000\,km to more than 10 Earth radii (1\,$R_\mathrm{E}\sim6400\,$km) \citep{zhang:2021} and they are thought to play an important role in the processes that enable the transfer of the SW mass, momentum and energy into the magnetosphere of our planet.

In this work we show that HFAs may replace FCBs at the edges of TFs. This occurs when the motional electric field points towards the IMF directional discontinuities bounding these foreshocks. We present two TFs that exhibit HFAs on one of their edges. These were observed by the constellation of the Cluster spacecraft on 5 January 2005 (Section~\ref{sec:20050105}) and 12 January 2005 (Section~\ref{sec:20050112}).

We also show that on occasions, FCBs bounding TFs may have an HFA-like appearance in the spacecraft data. However, a more detailed analysis reveals that the apparently enhanced temperatures and decreased and deflected velocities in their cores do not stem from heating of the SW ion beam. In fact, there is no evidence for the SW beam heating inside these events and the beam is strongly reduced inside them. The measured plasma moments are heavily influenced by the suprathermal ions. The latter tend to be hotter and more diffuse than their counterparts inside the corresponding TFs. We feature two HFA-like FCBs observed by the MMS mission on 6 February 2022 (section~\ref{sec:20220206}) and on 30 January 2022 (Section~\ref{sec:20220130}). 

In the following we describe the datasets used for the purpose of this study in Section~\ref{sec:data}, we feature the events in Sections~\ref{sec:caseHFA} and \ref{sec:caseHFA-like} and discuss our findings in Section~\ref{sec:discussion}. Finally, we present our conclusions in Section~\ref{sec:conclusions}.

\section{Instruments and Data Sets}
\label{sec:data}
In this work we use observations from two multi-spacecraft missions: Cluster \citep{escoubet:2001} and Magnetospheric Multiscale Mission \citep[MMS, ][]{burch:2016}. Both missions consist of four identical satellites that continuously measure magnetic field and plasma properties in the near-Earth environment.

The Cluster spacecraft carried several instrument suits. Here, we use the magnetic field data from the Fluxgate Magnetometer \citep[FGM, ][]{balogh:2001} with a time resolution of 0.2~seconds. We also examine plasma data with a time resolution of $\sim$4~seconds from the Hot Ion Analyser (HIA) instrument which forms part of the Cluster Ion Spectrometer \citep[CIS, ][]{reme:2001} package.

In the case of the MMS mission we use plasma data from the Fast Plasma Instrument \citep{pollock:2016} with a time resolution of 4.5 seconds and the magnetic field data from the Fluxgate Magnetometer \citep{russell:2016} with a time resolution of 4.5 second and 1/16th of a second.

\section{HFAs at the edges of traveling foreshocks}
\label{sec:caseHFA}
\subsection{5 January 2005 event}
\label{sec:20050105}
\begin{figure}
\centering
\includegraphics[width=1.1\textwidth]{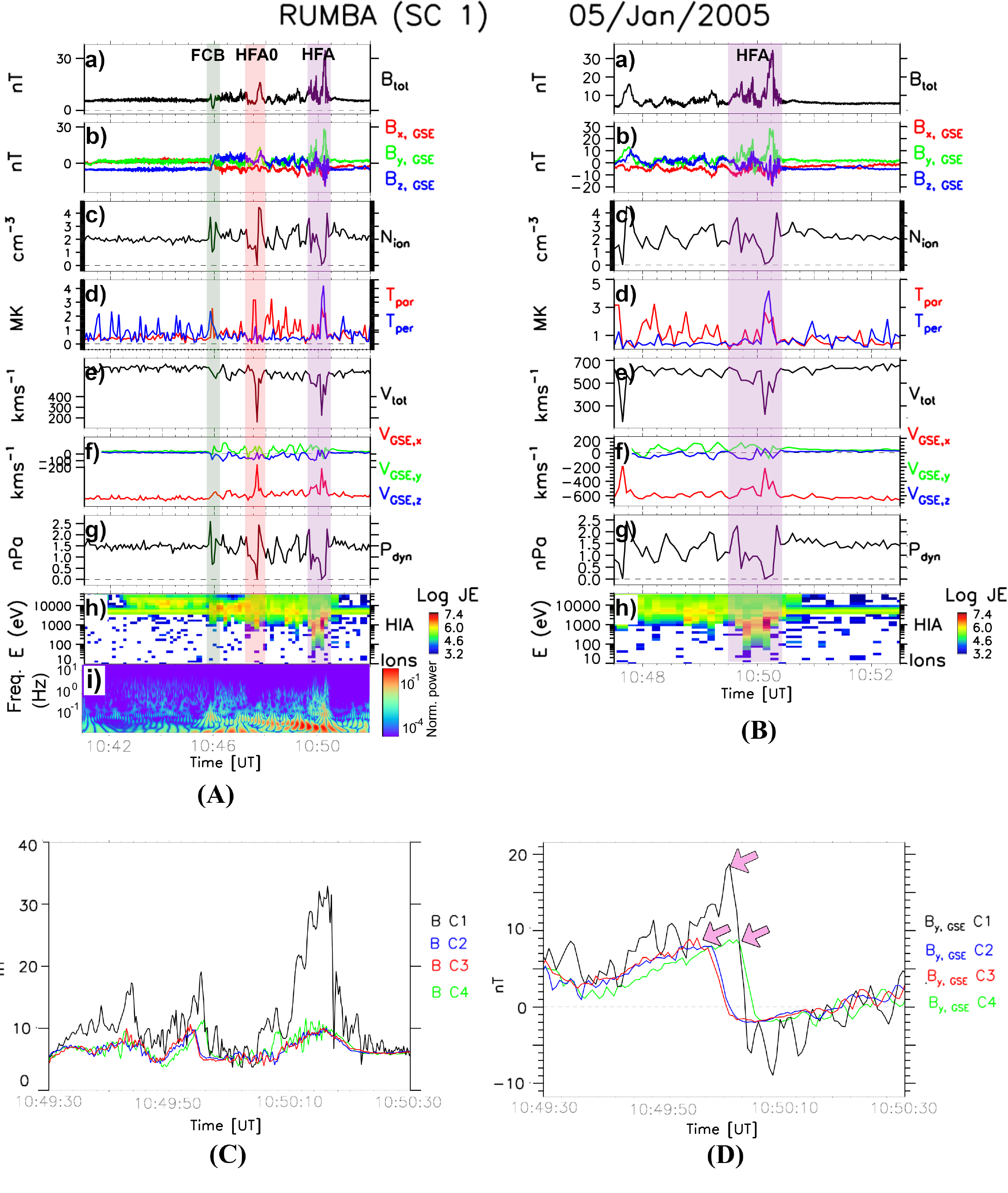}
\caption{(A) Traveling foreshock observed on 5 January 2005 by the Cluster 1 spacecraft. Two events with major solar wind flow deceleration are associated to it. The first is a SHFA which was detected at 10:47~UT deep inside the traveling foreshock (red shading). The second event is an HFA detected at $\sim$10:50~UT at the edge of the traveling foreshock (purple shading). (B) A shorter time interval centered on the HFA. Panels in Figures (A) and (B) exhibit (from top to bottom) IMF magnitude, IMF components in GSE coordinate system, plasma density, perpendicular (blue) and parallel (red) temperatures, SW bulk speed, SW velocity components in GSE coordinate system, dynamic pressure and ion spectrum. Additionally, panel Ai exhibits a Morlet wavelet spectrum of the B$_{X,GSE}$ component. (C) Magnetic field profiles of the four Cluster spaceceraft during the HFA. (D) B$_y$ profiles of the IMF directional discontinuity at all spacecraft. The arrows mark the times used by the timing analysis.}
\label{fig:20050105}
\end{figure}

The first event was observed on 5 January 2005 by the Cluster 1 (C1) spacecraft. Figure~\ref{fig:20050105}A exhibits an 11-minute time interval during which the TF with a HFA was observed. At $\sim$10:42~UT, C1 entered the foreshock region marked by enhanced fluxes of suprathermal ions (orange, green and yellow colors with energies above $\sim$8~keV on panel h). There were no 30~second ($\sim$0.03~Hz) ULF waves were present at that time (see panels a, b and the wavelet spectrum of the B$_{X,GSE}$ component on panel i), indicating that the observed suprathermal ions were FABs \citep[e.g., ][]{meziane:2005}, which we show explicitly below. C1 did not detect any IMF directional discontinuity upon entering this region, which thus probably belongs to the ``normal'' foreshock of Earth.

At 10:45:57~UT C1 detected an IMF directional discontinuity that was followed by a region populated by compressive ULF fluctuations and different types of ion populations. Among the latter are the SW beam ions that have not been heated and are visible on the panel h as a continuous narrow yellow strip that is present throughout the entire exhibited time interval, suprathermal foreshock ions with energies above that of the SW beam that are present throughout the foreshock interval and locally heated SW ions with a much wider range of energies that also span below that of the SW beam. Concurrently with the discontinuity, C1 observes an FCB marked by initial rise in B and n values, followed by their drop compared to surroundings.

At $\sim$10:47~UT the plasma velocity decreased from $\sim$640~km s$^{-1}$ to 145~km s$^{-1}$, mainly due to the changes in V$_X$ component (red shading in Figure~\ref{fig:20050105}A. The region of diminished SW speed occurs together with strong decreases in plasma density and the IMF magnitude and it is delimited by two rims with increased values of B and N. The parallel ion temperature (red trace panel d) there is enhanced from 1~MK to about 3~MK, while the perpendicular temperature exhibits no major variations. Due to these signatures and the fact that the magnetic field across it rotates by $\sim$30~$^\circ$, we denominate is as HFA0.

At the trailing edge of the traveling foreshock ($\sim$10:50~UT, purple shading) we observe another event presented in more detail in Figure~\ref{fig:20050105}B. It exhibits a set of signatures typical of HFAs: a core with strongly diminished plasma density (almost to $\sim$0) and highly fluctuating magnetic field that is bounded by two rims where both of these quantities are enhanced. In the core of this HFA, the parallel ion temperature (red trace) rises from $\sim$0.5~MK to $\sim$2.7~MK, while the perpendicular temperature (blue trace) jumps to $\sim$3.8~MK. The SW velocity drops from 640~km s$^{-1}$ to 215~km s$^{-1}$ inside the core of the event, mainly due to the V$_x$ component. Because of the strong drop in V and N, the dynamic pressure P$_{dyn}$ also diminishes. The IMF rotates across the associated IMF directional discontinuity by $\sim$37$^{\circ}$. The ion spectrum (panel h) indicates that the temperature rise inside the HFA core is largely due to the strong heating of the pristine SW ion population.

\begin{figure}
\centering
\includegraphics[width=0.8\textwidth]{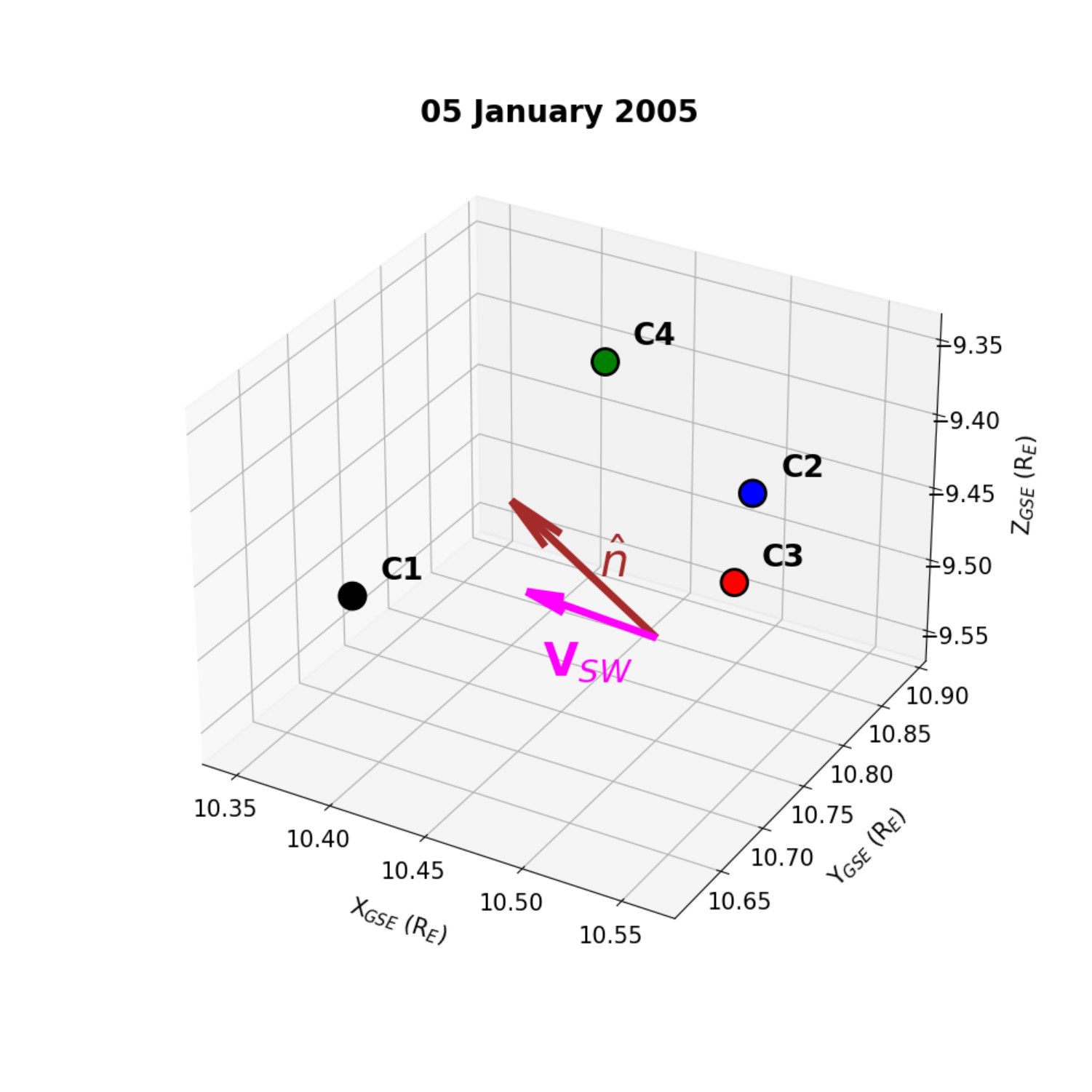}
\caption{Locations of the Cluster probes during the detection of the IMF directional discontinuity on 5 January 2005. The brown arrow represents the vector normal to the directional discontinuity while the magenta arrow indicates the direction of the SW velocity.}
\label{fig:pos20050105}
\end{figure}

In the Figure~\ref{fig:20050105}C we show the magnetic field profiles from all Cluster satellites. We can see that C2, C3 and C4 probes did not observe HFA signatures (enhanced rims, pronounced central dip). Additional inspection of the C3 plasma data (not shown) revealed lack of plasma heating indicating that these three probes probaly observed an FCB or, alternatively a very young HFA. The reason for this can be inferred from Figure~\ref{fig:pos20050105} where we show the locations of the Cluster probes during the detection of the HFA/FCB. At that time, the X$_{GSE}$ coordinate of the C1 spacecraft was the smallest of all, indicating that this spacecraft was closest to the bow shock. This was verified by calculating the distances of all four spacecraft from the model bow shock described by \cite{slavin:1981} as a hyperboloid with eccentricity of 1.15, the location of the foci at (3,0,0)~R$_E$ and semi-latus rectum of 23.75~R$_E$. The latter value was obtained by scaling the original \cite{slavin:1981} model by using the ambient SW dynamic pressure values, as described in \cite{jelinek:2012}. Thus, this mature HFA only stretched slightly beyond the C1 location but not all the way to the other probes.

All four spacecraft observed the directional discontinuity at the downstream edge of the traveling foreshock at slightly different times. We use the timing method similar to the one described by \cite{schwartz:1998} to calculate the normal to the plane of the discontinuity. In the Figure~\ref{fig:20050105}D we present the B$_y$ profiles from all Cluster satellites during the time interval that contains the directional discontinuity. In order to calculate the normal, we use the times of a common feature that appears in the data of all four probes: the moment when the values of the B$_y$ component start diminishing towards negative values (marked by arrows). The Figure~\ref{fig:20050105}D reveals that the order in which the probes detected this feature is C3 (red), C2 (blue), C1 (black) and C4 (green). The corresponding times were 10:49:55.461~UT, 10:49:54.517~UT, 10:49:54.265~UT and 10:49:55.902~UT, respectively. 
We calculate the normal vector to the discontinuity in the GSE coordinates to be $\hat{n} = (-0.81, 0.11, 0.57)$.

The direction of the normal is marked in the Figure~\ref{fig:pos20050105} with a brown arrow, while the SW speed is represented with the magenta arrow.

We proceed by evaluating the type of the discontinuity, i.e. tangential versus rotational. We do so by calculating the following values from the C1 data and adopting the criteria from \cite{neugebauer:1984}:
 \begin{linenomath*}
 \begin{eqnarray}
    B_n/B = 0.46,\\
    \Delta B/B = 0.024,
 \end{eqnarray}
 \end{linenomath*}
where $B_n$ is the component of the IMF normal to the discontinuity plane and the $\Delta B$ is the difference between the magnetic field magnitudes on both sides of the discontinuity. B is the magnitude that is equal to the larger of the two. According to the classification by \cite{neugebauer:1984} this discontinuity fulfills the criteria for a rotational discontinuity ($\Delta B/B < 0.2$ and $B_n/B \geq 0.4$).

We also check whether conditions for an HFA formation have been met by calculating the directions of the convection electric field  ($-\mathbf{V}\times\mathbf{B}$) on both sides of the discontinuity. We use the C1 data averaged over a $\sim$1-minute time intervals. 

Inside the traveling foreshock, the angle between the normal to the plane of the discontinuity and the convection electric field is equal to 106$^\circ$, while outside it is 79$^\circ$. Thus, the first condition for the formation of the HFA (convection electric field pointing towards the discontinuity on at least one side) is met.

\begin{figure}
\centering
\includegraphics[width=1.20\textwidth]{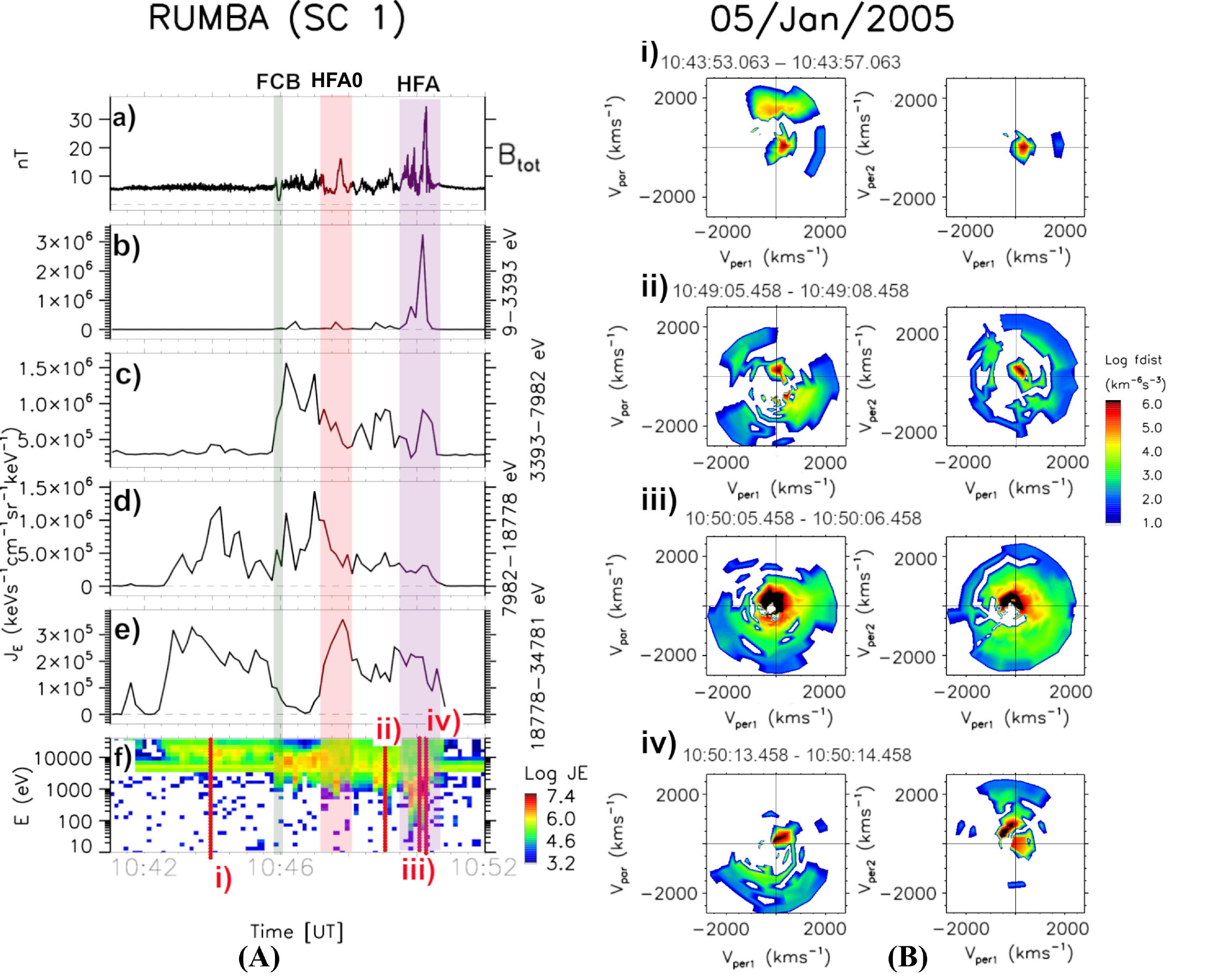} 
\caption{(A) From top to bottom: magnetic field magnitude, ion fluxes in different energy ranges and ion spectra during the time period when the traveling foreshock on 5 January 2005 was observed. (B) Ion distribution functions at selected times (marked in Figure~\ref{fig:flux20050105}Af). }
\label{fig:flux20050105}
\end{figure}

We further analyze ion fluxes and velocity distribution functions (VDFs) during times when the traveling foreshock was observed. The Figure~\ref{fig:flux20050105}A shows magnetic field magnitude (a), ion spectra (f) and ion energy fluxes in four different energy ranges that correspond to ions with energies below the SW beam energy (b), the SW beam energies (c) and to suprathermal energy ranges (d and e). Vertical lines on the bottom panel indicate the times of the ion VDFs exhibited in Figure~\ref{fig:flux20050105}B.

The most notorious feature in the Figure~\ref{fig:flux20050105}A is the strong peak in the flux of ions with energies below that of the SW inside the HFA. The flux also peaks in the SW beam energy range (panel c), although the peak is less prominent compared to the rest of the TF interval. Ions in higher energy ranges (panels d and e) do not display any special behavior.

Figure~\ref{fig:flux20050105}B shows ion distributions inside the traveling foreshock and in its vicinity in the SW bulk rest frame. The left panels feature VDFs projected onto the ($V_{per1}$, $V_{par}$) plane, where $V_{par}$ is the velocity component parallel to the background IMF, while ${V}_{per1}$ direction is calculated as $-\bf{V}\times\bf{B}$, so it is parallel to the direction of the convection electric field. The right panels show VDFs in the (${V}_{per1}$, ${V}_{per2}$) plane, where $\bf{V}_{per2} = \bf{V}_{par}\times\bf{V}_{per1}$ (direction of electric field drift velocity). The described planes are chosen so that they contain the origin.

Figure~\ref{fig:flux20050105}Bi shows that before the arrival of the TF, there are two dominant ion populations: one that constitutes the SW wind beam (close to the origin of the two panels) and the second one with positive (sunward) parallel velocities, thus forming a field-aligned ion beam (FAB). Figure~\ref{fig:flux20050105}Bii shows ion distributions just prior the HFA. The SW beam is still present, however there is a second population that forms a partial ring around the SW beam on the left panel of Figure~\ref{fig:flux20050105}Bii.The right panels show that the distribution is basically gyrotropic \citep{fuselier:1995}. Thus, this population can be described as in a stage between gyrating and diffuse. 

In the core of the HFA (\ref{fig:flux20050105}Biii), the SW beam and suprathermal VDFs cover a much wider range of velocities, revealing a strong heating of both populations. Finally, the Figure~\ref{fig:flux20050105}Biv shows the distribution inside the rear edge of the traveling foreshock. Besides very intense and anisotropic SW population, the VDF also exhibits the presence of suprathermal ions that form intermediate, gyrophase-bunched population \citep{fuselier:1995}. After that time, the ion VDF becomes that of the pristine SW beam (not shown).

\subsection{12 January 2005 event}
\label{sec:20050112}
\begin{figure}
\centering
\includegraphics[width=1.20\textwidth]{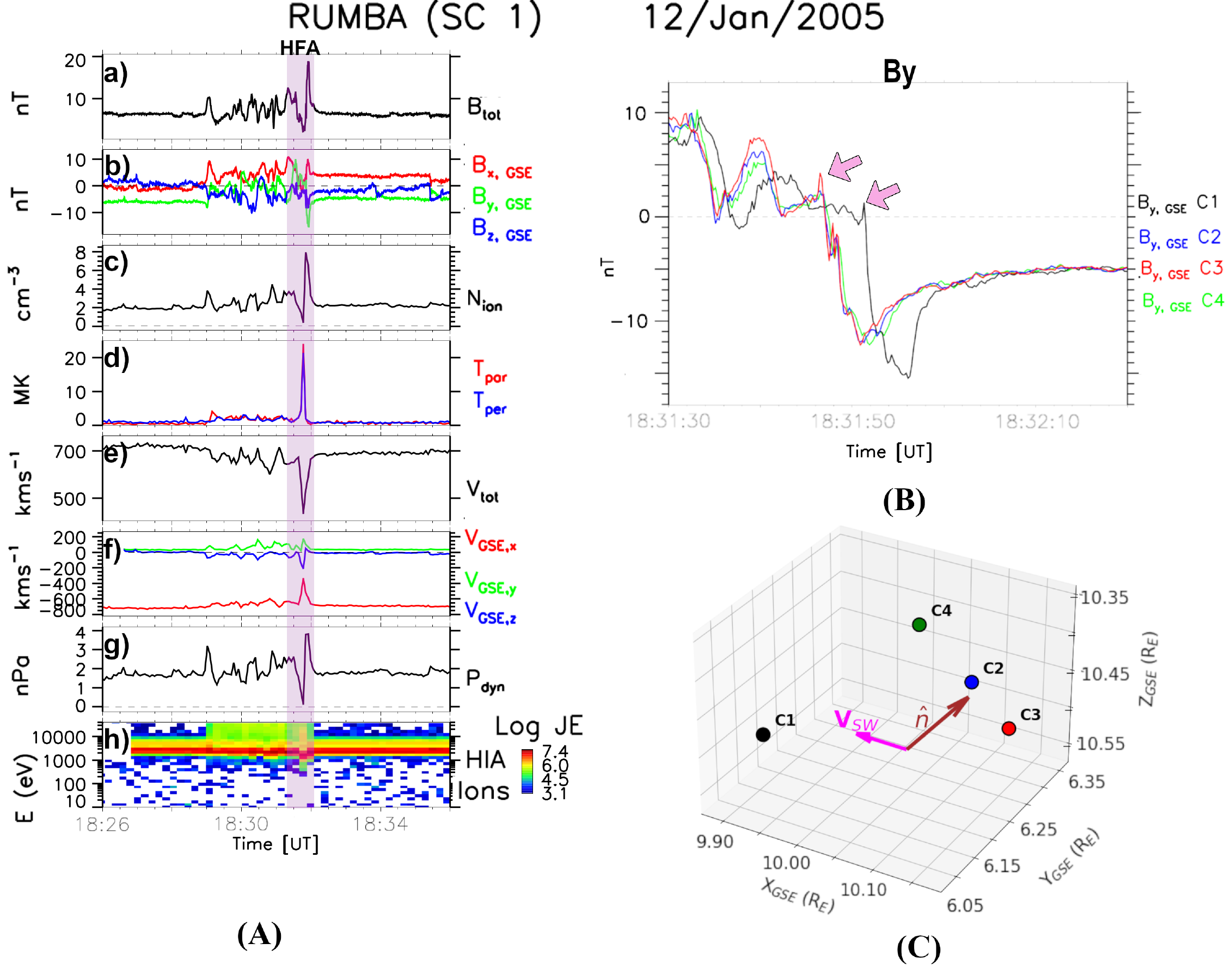} 
\caption{(A) Cluster 1 observations of a traveling foreshock with an HFA on its rear edge observed on 12 January 2005. (B) B$y$ profiles of the rear edge of the traveling foreshock. The arrows mark the times used in the timing analysis. (C) Positions of the Cluster probes during the detection of the IMF directional. The brown arrow represents the vector normal to the directional discontinuity while the magenta arrow indicates the direction of the SW velocity.}
\label{fig:20050112}
\end{figure}

Figure~\ref{fig:20050112}A shows the time interval containing a traveling foreshock observed on 12 january 2005. On its trailing edge there is an HFA. The B-magnitude inside its core diminishes to 1.6~nT and the plasma density almost to zero. The parallel and perpendicular temperatures increase from $\sim$0.8~MK to $>$23~MK and the SW flow decelerates from 660~km s$^{-1}$ to 420~km s$^{-1}$. This is mainly due to the change of the V$_x$ component, although the other two also exhibit significant change inside the HFA. The ion spectrum also shows ion heating in the core of the HFA. This time, all four Cluster probes observed the HFA.

Figure~\ref{fig:20050112}B exhibits B$_y$ profiles during the second directional discontinuity from the four Cluster probes. We can see that the order in which the spacecraft detected it was C3, C4, C2 and C1. The time delays with respect to the C1 satellite were -4.8~s, -4.76~s and -4.65~s for C2, C3 and C4, respectively. Figure~\ref{fig:20050112}C exhibits the locations of the four Cluster probes at the time when they detected the HFA.

From the timing method we obtain the normal to the plane of the discontinuity to be $\hat{n} = (-0.51, -0.64, -0.57)$.

The angles between the normal to the discontinuity plane and the convection electric field inside and outside the traveling foreshock are 44$^\circ$ and 76$^\circ$, which means that the HFA formation condition is met.
Again, we calculate the quantities $B_n/B = 0.15$ and $\Delta B/B = 0.12$.

According to \cite{neugebauer:1984}, the true character of such a discontinuity cannot be unambiguously determined, and the discontinuity could be either rotational or tangential. 

\begin{figure}
\centering
\includegraphics[width=1.2\textwidth]{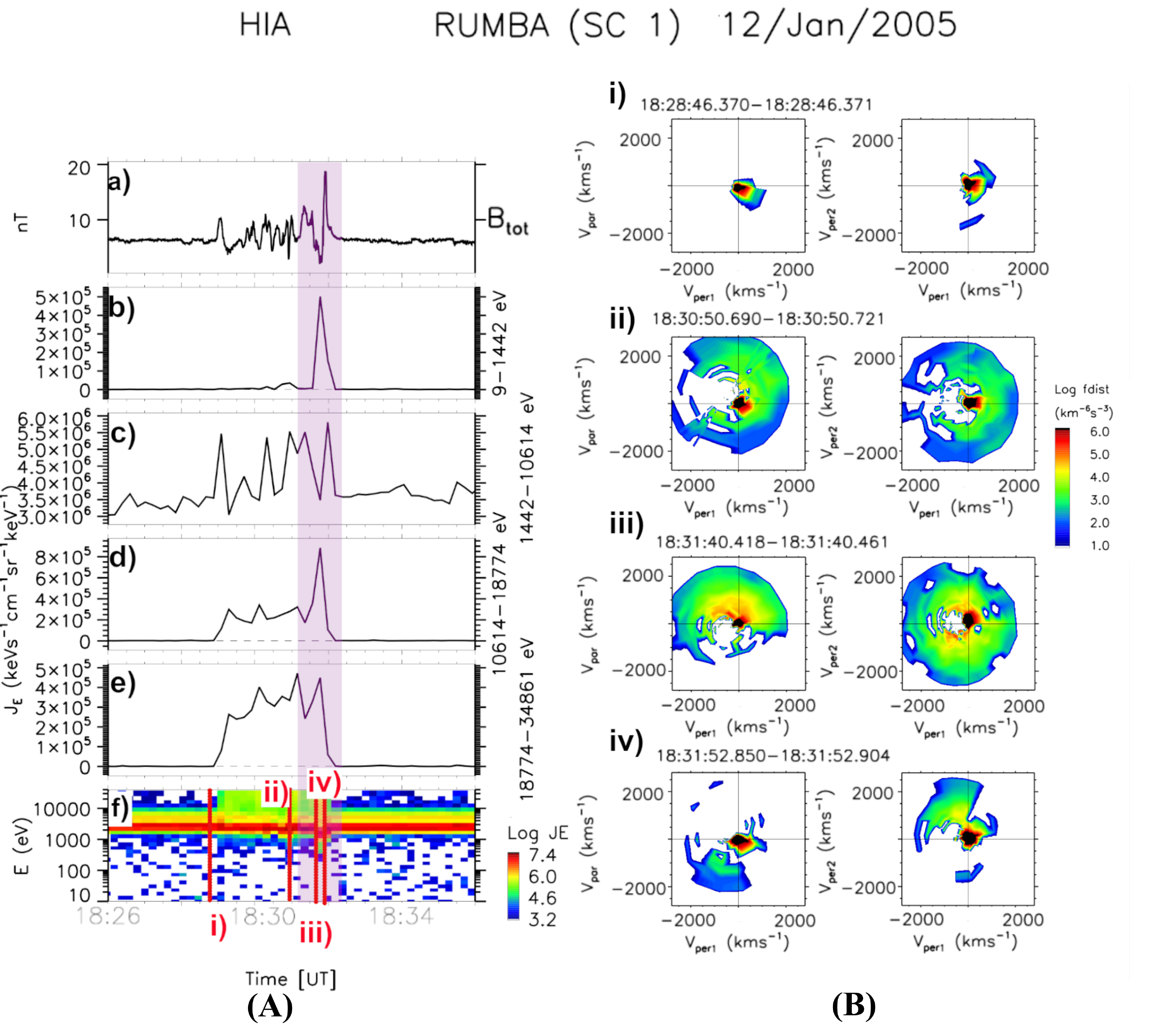}
\caption{(A) Ion fluxes around the time of the 12 January 2005 event. (B) Ion distribution functions at selected times (marked in Figure~\ref{fig:flux20050112}Af).}
\label{fig:flux20050112}
\end{figure}

We show ion fluxes in Figure~\ref{fig:flux20050112}A. It is clear that the fluxes below the SW energies (panel b) are strongly enhanced inside the HFA. Fluxes in the SW beam energy range (c) diminish compared to those just before the HFA. Suparthermal ions show a distinct peak just above the SW energies (d) and a less prominent one in the highest energy range (e).

In Figure~\ref{fig:flux20050112}B we show ion distributions at four selected times. Panel i) shows ion VDFs before the arrival of the traveling foreshock when only the SW ions are present. Later, well inside the foreshock (panel ii), there exists another, an almost completely diffuse ion population. Inside the HFA core (panel iii), the suprathermal ions exhibit intermediate-type of distribution, their VDF also appears more intense and hotter than inside the TF. Inside the rear rim of the HFA (panel iv) the ions form a gyrophase-bunched population \citep{fuselier:1995}.

\section{HFA-like foreshock compressional boundaries}
\label{sec:caseHFA-like}
In this section we present two events that exhibit many of the observational signatures of HFAs, however a careful inspection reveals the absence of the SW beam heating in their cores.

\subsection{6 February 2022 event}
\label{sec:20220206}
\begin{figure}
\centering
\includegraphics[width=1.2\textwidth]{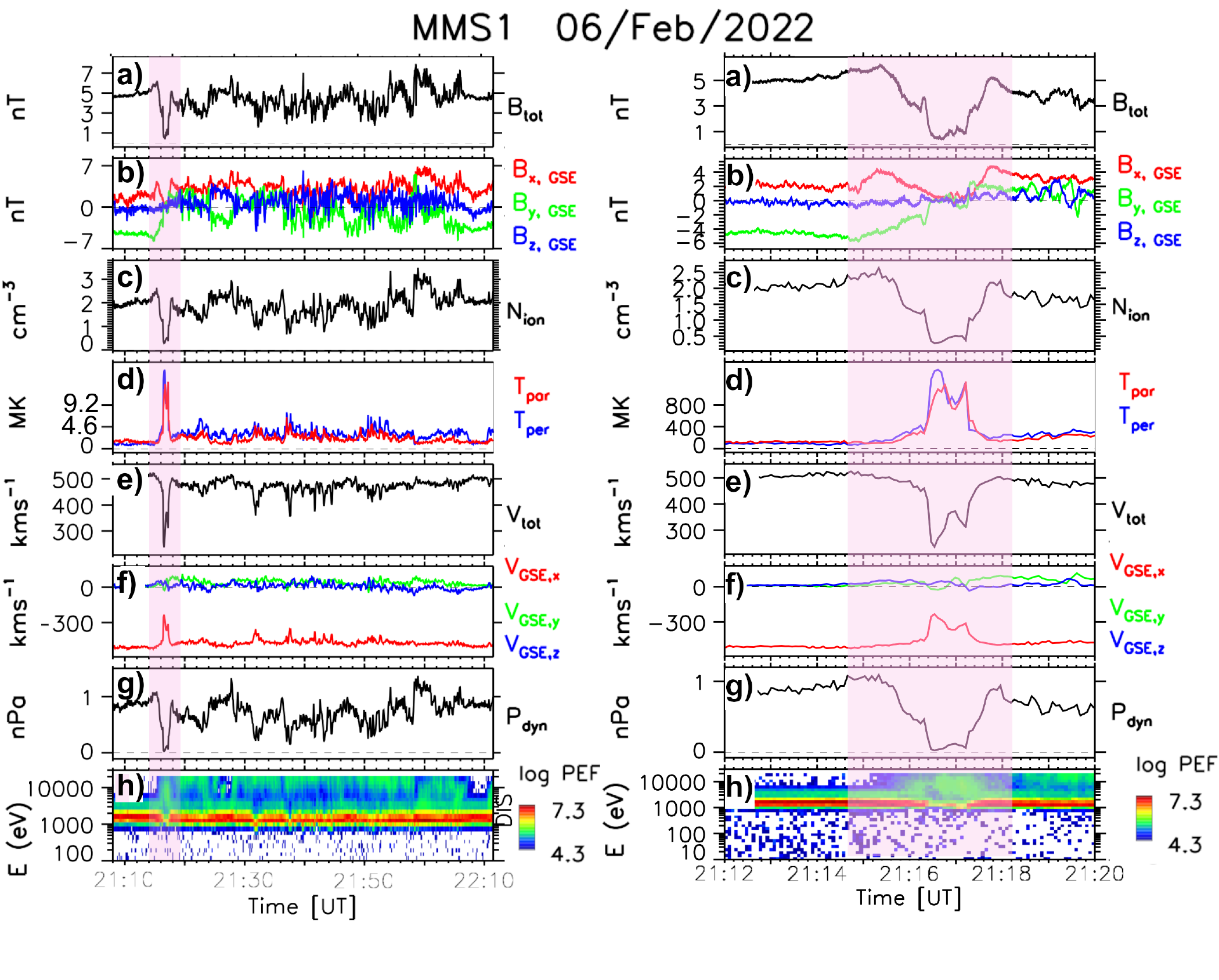}
\caption{Traveling foreshock and HFA-like event observed on 6 February 2022.}
\label{fig:20220206}
\end{figure}

Our third case study consists of an event that was observed by the MMS spacecraft on 6 February 2022. At first glance, its signatures in Figure~\ref{fig:20220206}A and B) are similar to the HFAs: it appears at the downstream edge of a traveling foreshock and is centered on an IMF directional discontinuity. It exhibits strong temperature rise in the core of the event (to $\sim$16.5~MK) and the flow is significantly deflected and decelerated from $\sim$500~km s$^{-1}$ to $\sim$240~km s$^{-1}$. The minimum B value inside the structure is only $\sim$0.16~nT and the density falls to $\sim$0.24~cm$^{-3}$.

This event, however, is different from the previous ones. The detailed inspection of the ion spectra in Figure~\ref{fig:20220206}Bh reveals that the SW wind beam does not undergo any heating at all. Rather, its intensity is strongly diminished and suprathermal ions are dominant in the core of the event.

\begin{figure}
\centering
\includegraphics[width=1.2\textwidth]{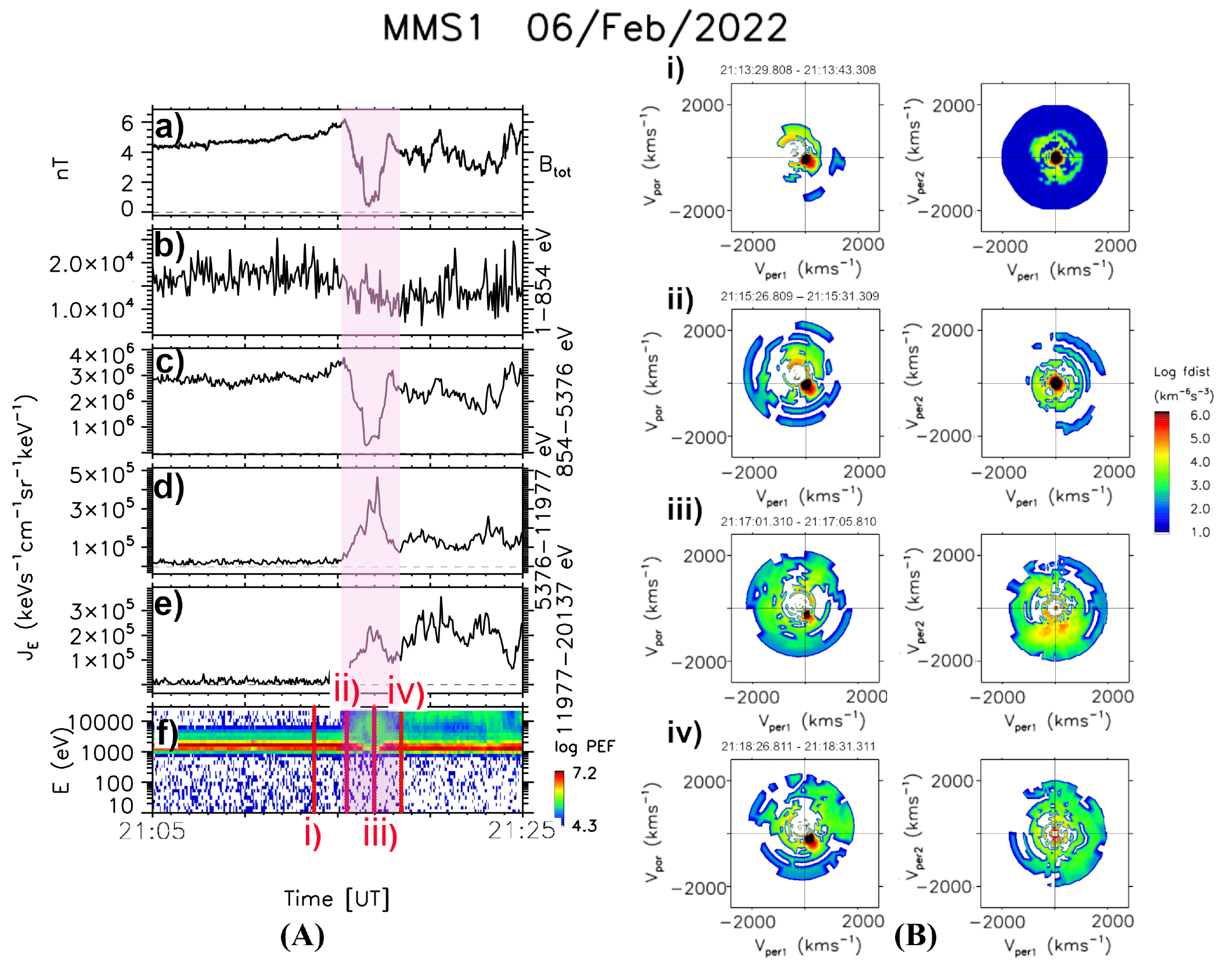}
\caption{(A) Magnetic field and ion fluxes during the event observed on 6 February 2022. (B) on distribution functions at selected times marked in \ref{fig:flux20220206}Af).}
\label{fig:flux20220206}
\end{figure}

We present ion energy fluxes in Figure~\ref{fig:flux20220206}A. We can see that the ion fluxes in the energy range below the pristine SW beam inside the structure are similar to those in the traveling foreshock, while in the energy range of the SW beam they diminish greatly. Suprathermal ions energy fluxes exhibit peaks inside the event.

We exhibit the ion distributions associated with this event in Figure~\ref{fig:flux20220206}B. Their times are indicated in Figure~\ref{fig:flux20220206}Af. Before the arrival of the traveling foreshock (panel i), there seem to be two ion populations present. The detailed inspection of the ion spectra and of the moments provided by Hot Plasma Composition Analyzer \citep[HPCA, ][not shown]{young:2016} reveal that both populations belong to the pristine SW beam. The one with $\sim$zero velocity in Figure~\ref{fig:flux20220206}Bi is formed by protons, while the other one by alpha (He$^{++}$) particles.

At the front rim of the event (panel ii) the two SW populations are joined by a hotter population which appears detached from them. In the core of the event (panel iii) the suprathermal population is hotter and almost completely diffuse, while the SW beam intensity is strongly diminished. After the event (panel iv), the suprathermal population is almost completely diffuse however colder than the one inside the event.

\subsection{30 January 2022 event}
\label{sec:20220130}
\begin{figure}
\centering
\includegraphics[width=1.2\textwidth]{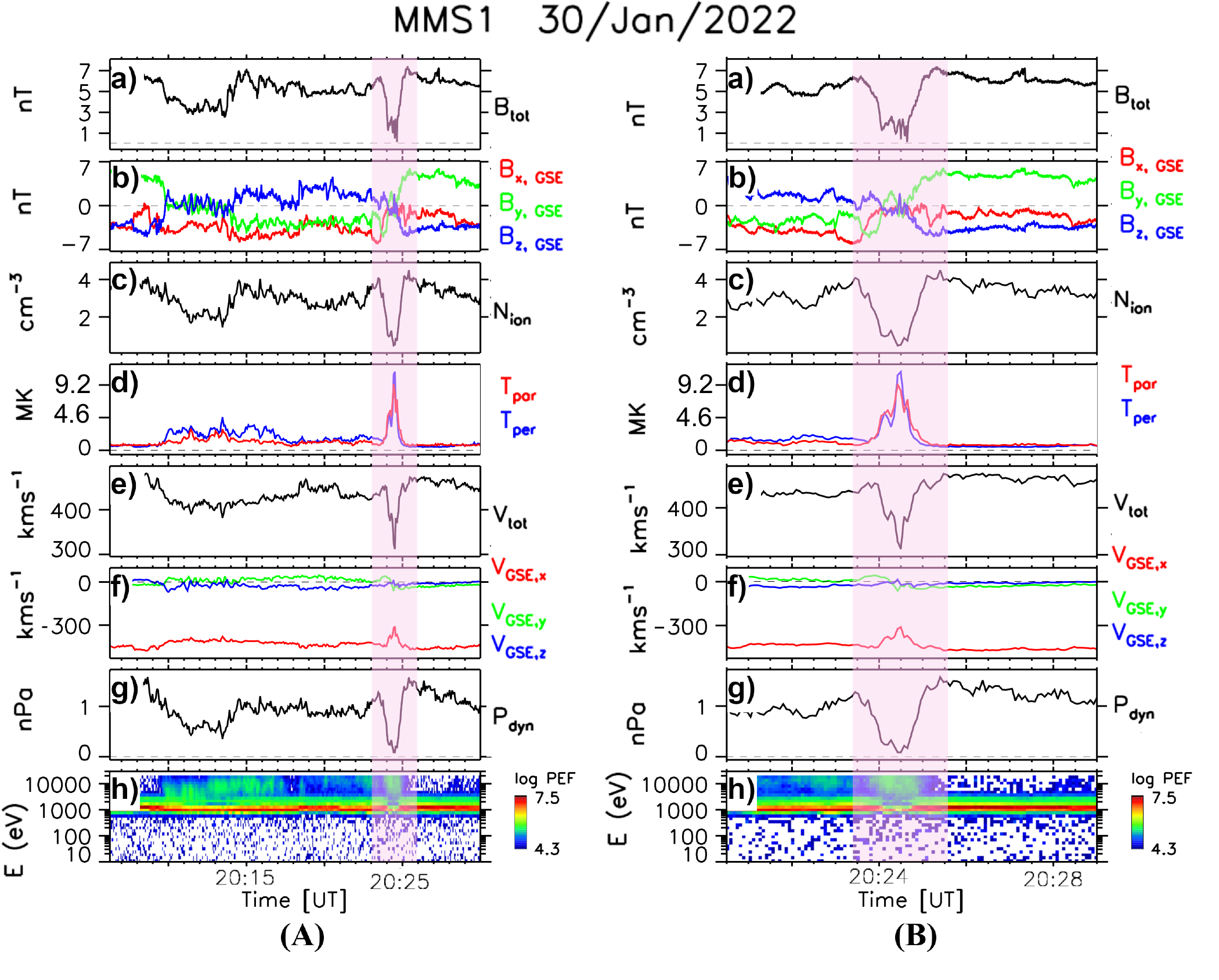}
\caption{Traveling foreshock and HFA-like event observed on 30 January 2022.}
\label{fig:20220130}
\end{figure}

Our last case study was observed on 30 January 2022 by the MMS mission (Figures~\ref{fig:20220130}A and B). The corresponding traveling foreshock was first detected at $\sim$20:10~UT and it lasted until $\sim$20:25~UT. The HFA-like event on its rear edge lasted from $\sim$20:23:45~UT until 20:25:37~UT. The lowest B value inside it is almost 0~nT, while the lowest density value is 0.35~cm$^{-3}$. The temperature inside it peaks at 10.6~MK, while the SW speed drops from 433~km s$^{-1}$ inside the foreshock to the minimum value of 310~km s$^{-1}$ inside the event. 

\begin{figure}
\centering
\includegraphics[width=1.2\textwidth]{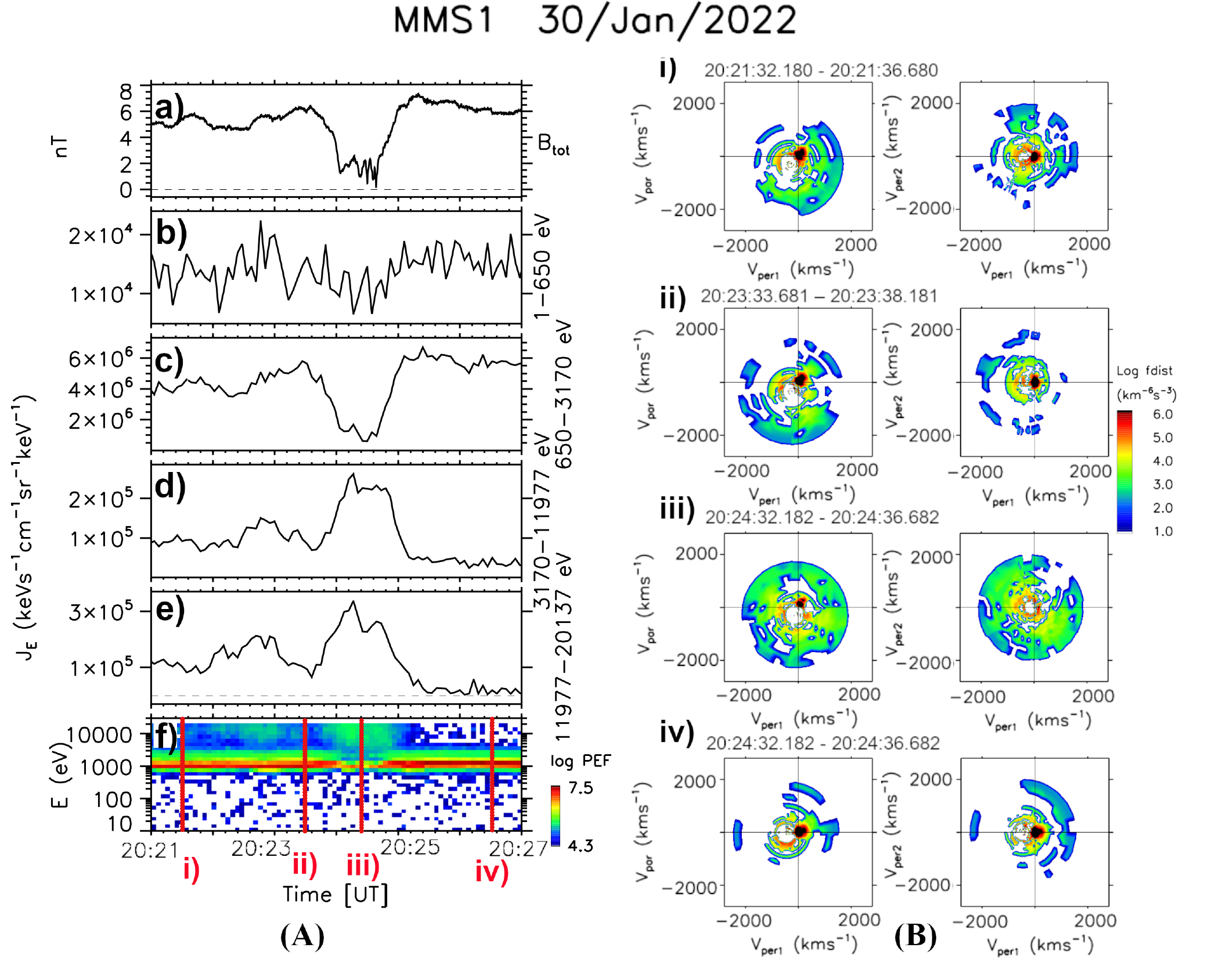}
\caption{(A) Magnetic field and ion fluxes during the event observed on 6 February 2022. (B) on distribution functions at selected times marked in \ref{fig:flux20220130}Af).}
\label{fig:flux20220130}
\end{figure}

We again analyze the ion fluxes in and in the vicinity of this structure in Figure~\ref{fig:flux20220130}A. We see that the flux of ions with energies below that of the SW (b) exhibit similar levels to those inside the traveling foreshock. The fluxes in the SW beam range (c) decrease strongly inside the event, while the suprathermal ion fluxes (d and e) exhibit distinct peaks.

From Figure~\ref{fig:flux20220130}B we also see that the SW ion population significantly decreases in the core of the structure (\ref{fig:flux20220130}Biii). The VDFs of suprathermal ions change from intermediate before the event (\ref{fig:flux20220130}Bi) and in the front rim of the structure (\ref{fig:flux20220130}Bii), to a hot, diffuse population in the core of the event (\ref{fig:flux20220130}Biii). At later times (\ref{fig:flux20220130}B)iv) the pristine SW beam is dominant.

\section{Discussion}
\label{sec:discussion}
\subsection{Multi-TUMS events}
\label{subsec:discussionA}
It is already known that foreshocks, the ``regular'' and traveling, are commonly delimited by FCBs. However, TFs differ from the regular foreshock in that they form inside magnetic field flux tubes that connect to the nominally quasi-perpendicular shock in a quasi-parallel manner. Thus, in the spacecraft data, TFs are always bounded by IMF directional discontinuities. It seems natural that when the conditions are right, HFAs will appear instead of FCBs. 

Such HFAs will, however, replace FCBs only along the section of the directional discontinuity close to the bow shock. This is because they will start to grow in the vicinity of the shock and will only gradually expand towards upstream. This might have been the case for the 5 January 2005 HFA that was observed only by the C1 spacecraft, while the three more distant probes detected an FCB.

Additionally, at the front edge of this TF the Cluster spacecraft observed an HFA-like FCB, while in its interior an HFA was detected. This means that four different TUMS (TF, HFA-like FCB, HFA and the trailing FCB) were impacting the bow shock almost simultaneously. The impact of such a multi-TUMS event will be a subject of future investigations.

Such studies are important, since there exist only a handful of investigations that deal with multi-TUMS events. \cite{omidi:2020b}, \cite{omidi:2020b} and \cite{turc:2025} performed 3- and 2.5-D hybrid simulations in order to show that a single IMF discontinuity can result in generation of FBs and HFAs. In the simulations by \cite{omidi:2020} both of these structures impact the magnetosheath of Venus resulting in the sunward expansion of the ionosphere and outflow of planetary ions with FB having a more global impact. In the simulations by \cite{turc:2025} a single IMF directional discontinuity first drove an FB and later, upon intersecting the simulated Earth's bow shock, it also lead to the formation of multiple HFAs. The combined impact of such multiple event on the bow shock and the downstream regions has not yet been investigated but \cite{turc:2025} suggested that it is likely more intense than the impact of of each individual structure.
% Additionally, there is only one observational study performed by \cite{rojascastillo:2025} that shows that SHFAs can form inside traveling foreshocks.

\subsection{HFA-like FCBs at the edges of traveling foreshocks}
\label{subsec:discussionB}
In the second part of our work we describe two TFs that present HFA-like FCBs at one of their edges. In the spacecraft data these events exhibit signatures similar to HFAs: a strong decrease of B and plasma density inside their cores that are accompanied by decelerated and deflected SW flows and a strong increase of temperature. However, a more detailed inspection reveals that in reality, the SW ions are not heated inside these events. The apparent temperature rise occurs mainly due to the fact that the SW ions there are almost absent and the plasma moments are strongly influenced by the suprathermal ion population. The latter, however, is hotter and more diffuse than the suprathermal ions in the rest of the TFs, resulting in a higher temperature.

Additionally, both HFA-like FCBs exhibit relatively modest rims, with small increases of B and N, and their profiles are quite ``clean'', i.e. without large-amplitude B and N oscillations or waves inside them.

It is not clear at this point whether HFA-like FCBs are a new phenomenon that does not form due to explosive expansion, thus explainig the lack of active ion heating in their cores and weak rims, or whether they represent late phases of HFA evolution, whent the SW beam has been completely energized and converted into the suprathermal ion population and the expansion had slowed down and the rims had weakened.

An important difference between the IMF directional discontinuities associated with our first two case studies (the HFAs) and the last two events (HFA-like FCBs) is their duration in the spacecraft data. 
These were 38 seconds and 24 seconds for the 5 January 2005 and the 12 January 2005 events, respectively, and approximately 3 minutes and 15 seconds for the other two events.

We determined the thicknesses of the latter IMF directional discontinuities from their durations in the data and by calculating their normals using the cross product of the magnetic field vectors inside the TFs and in the pristine SW. The thicknesses were found to be 1.9~R$_E$ and 1.6~R$_E$ for the 6 February 2022 and the 30 January 2022 events, respectively. This is much more than the estimated Larmour radii of suprathermal ions in the 10~keV--30~keV energy range for both events that span between 0.37~R$_E$ and 0.75~R$_E$. Thus, it is safe to say that the two IMF rotations were so thick that the suprathermal ions have remained magnetized inside them.

This may be one of the reasons for why FB or HFA did not evolve at the edges of these TFs. \cite{archer:2015} and \cite{an:2020} suggest that in order for these structures to form, the IMF rotations need to occur on spatial scales shorter than the gyro-radius of suprathermal ions. Thus, we turn to the work of \cite{omidi:2013} in order to try to find an explanation for the observed HFA-like FCBs.

\cite{omidi:2013} use a 2.5-D (2-D in space and 3-D in currents and electromagnetic fields) global hybrid code in order to study the dynamics of FCBs. The authors performed several runs under steady and time-varying IMF directions. In their second run, \cite{omidi:2013} simulate FCBs delimiting a traveling foreshock (see their Figure 6). The FCB that forms at the TF's leading edge exhibits similarities with our observed events: dips in B and N profiles are much deeper than the amplitudes of the ULF waves and the temperature in the core of that structure is higher than in the rest of the TF. 

\begin{figure}
\centering
\includegraphics[width=1.2\textwidth]{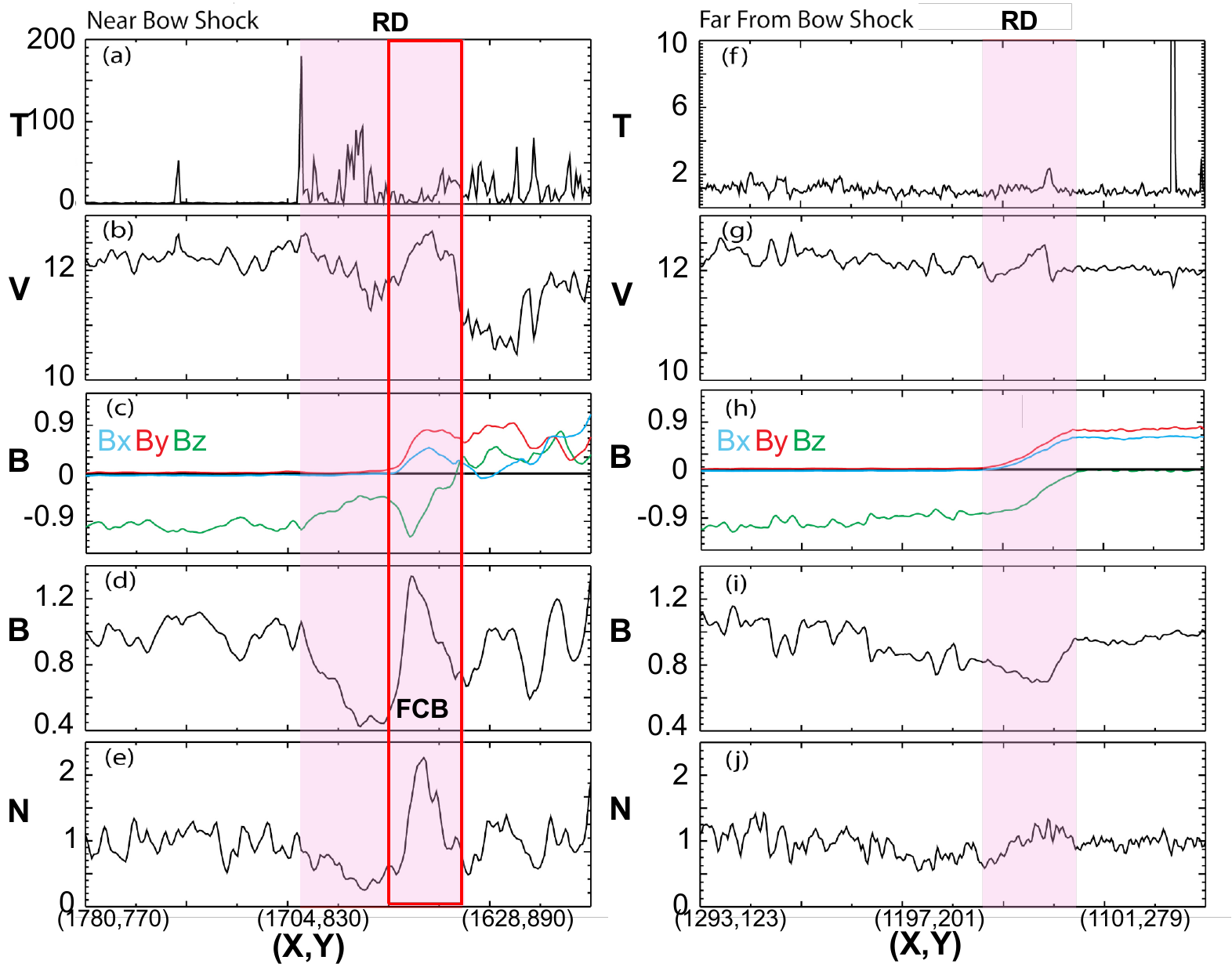}
\caption{Simulations results from \citep{omidi:2013}. Panels show profiles of (from top to bottom) temperature, ion velocity, IMF magnitude and components and plasma number density close to the bow shock (left column) and farther away from the simulated bow shock (right column).}
\label{fig:nick}
\end{figure}

To elaborate further, we show additional results from that run in Figure~\ref{fig:nick}. Panels (a)-(e) exhibit respectively, variations of temperature (normalized to upstream value), ion velocity (normalized to Alfven speed), and magnetic field components and magnitude and density (normalized to upstream values) along a trajectory near the bow shock. Starting in the SW, one encounters the RD (purple shading) which is associated with an increase in temperature, small drop in velocity and large decreases in magnetic field and density. The rear part of the RD coincides with enhancements in magnetic field and density which are associated with the formation of the FCB (encased within the red rectangle). As is evident in Figure 6 of \cite{omidi:2013}, the strength of the density enhancement associated with FCB decreases with distance from the bow shock (see panels (i) and (j) of Figure~\ref{fig:nick}). 

In general, the formation and properties of the FCB are not expected to be tied to the properties of the RD but related to the plasma conditions noted in \cite{omidi:2009, omidi:2013} such as the SW Mach number and the cone angle. Accordingly, we examine the properties of the RD by itself. In regards to the initialization of the RDs in the simulation, rotations in the magnetic field are implemented with a thickness of 2 thermal proton inertial lengths without any changes in plasma properties. With time, the thickness of RDs adjusts to reach a self-consistent structure which does not look like that seen in panels (a-e) in Figure~\ref{fig:nick}. To illustrate this point, panels (f-j) in Figure~\ref{fig:nick} show variations across the same RD as in panels (a)-(e) but at large distances from the bow shock. In contrast to the changes seen near the bow shock, except for some broadening and small drops in B within the RD no other changes are observed.

\section{Conclusions}
\label{sec:conclusions}
In this paper, we have examined the structure of the boundaries between TFs and the solar wind. Given that TFs are bounded by IMF RDs on both sides, the boundaries must include RDs as part of their structure. Past simulations and observations have shown that the edges of TFs may be associated with FCBs or FBs. In this study we show that TFs may also be associated with HFAs or nonlinear structures associated with the interaction of the RDs with the backstreaming ions from the bow shock. Past investigations have shown that formation of FCBs, HFAs and FBs all require certain plasma conditions which may or may not be met at the edges of a given TF. Accordingly, the edges of TFs, may be associated with any of these structures depending on the upstream conditions. We have also demonstrated that the interaction of RDs and backstreaming ions results in major changes in the properties of the RD including substantial decreases in density and magnetic field strength. This in turn impacts the overall structure of the TFs. Future work will show what is the impact if such structures on the bow shock and the regions downstream of it.

\section{Acknowledgments}
TP.K.'s work was funded by the PAPIIT-DGAPA through the project grant IN100424. DRC’s work was supported by the UNAM-PAPIIT IA105223 grant and SECIHTI CF-2023-I-1264 grant. N.O.'s work at Solana Scientific Inc. was supported by the NASA-LWS grant \#80NSCC23K1401. XBC acknowledges DGAPA PAPIIT grant IN106724.

\newpage
%TC:ignore

\newpage

\bibliography{References}
% \begin{thebibliography}{99}
% \end{thebibliography}
\end{document}